\documentclass[runningheads]{llncs}
\usepackage[T1]{fontenc}
\usepackage{graphicx}
\usepackage{color}
\usepackage{hyperref} 

\usepackage[ruled, vlined]{algorithm2e}
\usepackage{amsmath}
\usepackage{amssymb}
\usepackage{xcolor}
\usepackage{hyperref}

\usepackage{adjustbox}
\usepackage{pgfplotstable}
\pgfplotsset{ /pgfplots/ybar legend/.style={ /pgfplots/legend image code/.code={ \draw[##1,/tikz/.cd,bar width=3pt,yshift=-0.2em,bar shift=0pt] plot coordinates {(2*\pgfplotbarwidth,0.6em)};},}}
\pgfplotsset{compat=1.18}

\def\axisdefaultheight{0.47\textwidth}
\pgfplotsset{every axis/.style={scale only axis}}

\newcommand{\bigO}[1]{\mathcal{O}\left(#1\right)}

\newcommand{\users}{\mathcal{U}}
\newcommand{\facs}{\mathcal{F}} 
\newcommand{\bfx}{\mathbf{x}}
\newcommand{\bfX}{\mathbf{X}}
\newcommand{\bfy}{\mathbf{y}}

\newcommand{\bfA}{\mathbf{A}}

\newcommand{\bfr}{\mathbf{r}}
\newcommand{\bfc}{\mathbf{c}}
\newcommand{\bfa}{\mathbf{a}}

\begin{document}
%

\title{Bounding the Price-of-Fair-Sharing using Knapsack-Cover Constraints to guide Near-Optimal Cost-Recovery Algorithms
}


%
\titlerunning{Bounding the Price-of-Fair-Sharing}
%

\author{Sander Aarts
\and Jacob Dentes
\and Manxi Wu
\and David Shmoys
}
\authorrunning{S. Aarts et al.}
%
\institute{School of Operations Research and Information Engineering, Cornell University, Ithaca, NY}
\maketitle              
\begin{abstract}
We consider the problem of fairly allocating the cost of providing a service among a set of users, where the service cost is formulated by an NP-hard {\it covering integer program (CIP)}. The central issue is to determine a cost allocation to each user that, in total, recovers as much as possible of the actual cost while satisfying a stabilizing condition known as the {\it core property}. The ratio between the total service cost and the cost recovered from users has been studied previously, with seminal papers of Deng, Ibaraki, \& Nagomochi and Goemans \& Skutella linking this {\it price-of-fair-sharing} to the integrality gap of an associated LP relaxation. Motivated by an application of cost allocation for network design for LPWANs, an emerging IoT technology, we investigate a general class of CIPs and give the first non-trivial price-of-fair-sharing bounds by
using the natural LP relaxation strengthened with knapsack-cover inequalities. Furthermore, we demonstrate that these LP-based methods outperform previously known methods on an LPWAN-derived CIP data set.
We also obtain analogous results for a more general setting in which the service provider also gets to select the subset of users, and the mechanism to elicit users' private utilities should be group-strategyproof. The key to obtaining this result is a simplified and improved analysis for a cross-monotone cost-allocation mechanism. 

\end{abstract}

%
%

\section{Introduction}

Allocating costs fairly when providing some collective service to a set of users has been a mainstream topic within cooperative game theory in which the notion of the core, and the associated {\it core property}, plays a central role. Specifically, the core property requires that the payment from each user is such that each subset of users pays, in total, no more than the cost of providing service just to them. LP duality has long been known to play a fundamental role in such allocations, and the seminal work of Deng, Ibaraki \& Nagamochi \cite{DengAlgoAspectsCore} and Goemans \& Skutella \cite{goemans2004cooperative} proved the link between the integrality gap (or lack thereof) for several NP-hard discrete optimization settings and the fraction of the service cost that can be recovered by an allocation satisfying the core property. We term the ratio between optimal cost and the cost recoverable satisfying the core property as the {\it price-of-fair-sharing}.

For a natural subclass of {\it covering integer programs} (CIPs), known as the multiset-multi-covering problems, the strongest known price-of-fair-sharing was not obtained using a dual LP-based framework \cite{Li2005CovCostShare}. We show that an LP-based approach can yield improved bounds and simplified analyses, not just for this special case considered by \cite{Li2005CovCostShare}, but for general CIPs. More specifically, our main result is that by strengthening the natural LP relaxation for CIPs with knapsack-cover inequalities, the resulting dual LP does have the property that any feasible solution yields a cost-allocation that satisfies the core property. This reduces the problem of finding a cost-allocation to that of finding a strengthened-LP-relative approximation algorithm, allowing us to leverage approximation algorithms already known for these settings (e.g., \cite{carnes2008primal,carr1999strengthening,chekuri2019approximating,kolliopoulos2005}). Not only do these existing algorithms yield bounds on the price-of-fair sharing for general CIPs, they also provide existing bounds for families of sparse CIP instances.

We also obtain analogous results in a more general setting where the service provider selects a subset of users to receive the service based on their elicited private valuations of the service. The mechanism used to to elicit users' private valuations needs to be group-strategyproof so that no subset of users have the incentive to misreport their valuations. The key to obtaining this result is a simplified and improved analysis for a cross-monotone cost-allocation mechanism. These results also rely on our knapsack-cover-strengthened LP relaxation.
 
Our investigation into these questions was directly motivated by challenges in the development of an emerging technology, LPWANs, a popular Internet of Things solution for wirelessly connecting devices to the internet \cite{LoRa-technical}: CIPs emerge from the appropriate model for determining the location of gateways to provide coverage to a collection of users, and our result is useful to determine how the cost of LPWANs should be shared among users in an incentive compatible manner. We complement our algorithmic and structural results with an empirical investigation into the effectiveness of our methods on relatively large-scale LPWAN data. Our empirical study demonstrates that, in comparison to several natural heuristic alternatives, our theoretically-motivated methods provide far superior results.

\section{Optimizing coverage for LPWAN networks and IoT}
The growth of the Internet of Things (IoT) is creating both new opportunities and challenges in wireless coverage provision and pricing. 
Low-power wide-area networks (LPWANs) are a ubiquitous technology for connecting devices, or things, to the internet. These networks use radio communication to transmit signals over long distances. Demodulating, and transferring these signals to the internet requires networks of physical wireless receivers. Many such networks are already deployed.
In 2023 alone, LPWANs reportedly served over 1.3 billion IoT connections \cite{IoTAnalytics}. LoRaWAN is among the most popular LPWAN solutions, representing roughly 40\% of all connections made outside of China. Globally, over 150 different network operators provide LoRaWAN coverage \cite{LoRaEcosystems}.

Sharing LoRaWAN coverage among multiple users is an effective yet challenging approach to better utilize wireless infrastructure and lower costs. Two key features make sharing LoRaWAN coverage especially beneficial. Firstly, the receivers, or \textit{gateways}, can process orders of magnitude more wireless traffic than that produced by a typical single IoT application (see, e.g., the application in \cite{aarts2023interpretable}). Secondly, the quality of coverage improves as the number of gateways increases. Transmitted LoRaWAN packets tend to fail at random, and coverage quality is measured by the \textit{packet reception rate}. However, transmissions are \textit{association free}; a device broadcasts to all nearby gateways, many of which may process the same packet \cite{LoRa-technical}. A packet is lost if and only if all nearby gateways simultaneously fail to receive it. Thus, having multiple gateways to cover the transmitted signal can significantly improve the reception rate. This multi-coverage feature is a key aspect of our model and raises interesting computational challenges.

The problem of optimally providing wireless coverage to a group of users using multiple gateways can be modeled as a covering integer program. Let $\users$ denote a set of $m$ users, and $\facs$ a set of $n$ potential gateway locations, or facilities. Each user $j \in \users$ has a service requirement $r_j > 0$, and each gateway has an opening cost $c_i > 0$. Requirements and costs are represented by vectors $\bfr = (r_1, r_2, \dots, r_m)$ and $\bfc = (c_1, \dots, c_n)$, respectively. Each facility-user pair is associated with a contribution parameter $a_{ij} \geq 0$ that specifies the quality of coverage that facility $i$ provides user $j$ if opened; these can be estimated using data on the distance between the facility and the user as well as the terrain \cite{TorresSanzLoRaTerrain}. Contributions are represented as an $n \times m$ matrix $\bfA$. The objective is to minimize cost, subject to providing sufficient coverage to each user.
\begin{align}
\label{eq:integer_program}
c^* = \min\{\bfc^T \bfx: \bfA \bfx \geq \bfr, \  \bfx \in \{0, 1\}^n \}
\end{align}
Here, the binary decision variables $\bfx = (x_1, \dots, x_n)$ indicate whether each facility $i$ is opened or not. More details on how CIPs capture LoRaWAN coverage provision are given in Appendix \ref{sec:lorawan}. (A more general formulation of CIPs permits the purchase of up to $d_i$ integer copies of facility $i$. Our results extend to this setting in a straightforward way, but we omit this extension for simplicity and clarity of exposition.)

\section{Cost-sharing fundamentals}
The main objective in this work is to develop a principled algorithmic framework for finding fair and stable ways to share the cost of CIP solutions among the set of users. We follow a standard approach in mechanism design and define ``fair and stable'' using the \textit{core property} \cite{jain_mahdian_2007,parsons1998cross}. An allocation satisfies the core property if no subset of users is allocated more cost than the minimum cost of providing coverage for this subset alone. This property can also be viewed as an absence of cross-subsidies; no sub-group of users pays more than what it would cost them to serve themselves, thereby not subsidizing other users \cite{faulhaber1975cross}.

More formally, a cost allocation is a vector $\mathbf{\xi} = (\xi_1, \dots, \xi_m),$ where $\xi_j \geq 0$ represents the dollar amount charged to user $j \in \users$. Allocations $\xi_j$ are also called cost-shares.
Following the standard definition, cost-shares $\mathbf{\xi}$ satisfy the core property if for each sub-group $J \subseteq \users,$
\begin{equation}
    \label{eq:core_property}
    \sum_{j \in J}\xi_j \leq c^*_J, \quad \forall J \subseteq \users,
\end{equation}
where $c^*_J$ is the minimum cost of serving group $J$ \cite{jain_mahdian_2007}. Cost allocations can vary in magnitude. We say cost allocation $\mathbf{\xi}$ is \textit{budget-balanced} whenever the total amount paid by the served users, $\sum_{j \in \users}\xi_j$, equals the cost of serving them. The \textit{core} is the set of budget-balanced cost allocations satisfying the core property \cite{jain_mahdian_2007}. In some problems, finding cost-shares in the core can be challenging.

Linear programming (LP) duality is the workhorse tool for finding cost-shares satisfying the core property, but it is not always possible to simultaneously satisfy both the core property and to be budget-balanced. There are simple CIP instances that have no cost-allocations in the core (see, e.g., Li \textit{et al.} \cite{Li2005CovCostShare}). 
Deng, Ibaraki, and Nagamochi \cite{DengAlgoAspectsCore} show that a generic covering problem has a non-empty core if and only if its natural LP-relaxation has no integrality gap. As such, it is common to maximize the cost recovered, subject to satisfying the core-property \cite{goemans2004cooperative,jain_mahdian_2007};
a cost-allocation $\mathbf{\xi}$ is $\beta$-budget-balanced if it recovers a fraction $\beta$ of the cost.

For many optimization problems, every dual feasible solution is a cost allocation that satisfies the core property. 
This connection has important algorithmic implications. First, one can leverage duality to find cost-shares satisfying the core property by using (approximation) algorithms that produce feasible dual solutions as well as feasible integer solutions to the primal. If the integer solution costs at most $\alpha$ times the dual, we see that the corresponding cost allocation is $1/\alpha$-budget balanced. We shall say then that $\alpha$ is the {\it price-of-fair-sharing} (or equivalently, $1/\alpha$ has also been termed the cost-recovery ratio). This relationship has been frequently used to obtain strong results in many settings \cite{DengAlgoAspectsCore,devanur2003strategyproof,goemans2004cooperative,gupta2003approximation,konemann2008group,LeonardiShaefer2004,meir2010minimal,PalTardos2003group}.

On the other hand, seminal work by Deng, Ibaraki, \& Nagamochi \cite{DengAlgoAspectsCore} and Goemans \& Skutella \cite{goemans2004cooperative} shows that all cost allocations satisfying the core property are dual feasible solutions in the set cover problem and the facility location problem. Since the cost-allocation mechanism purchases an integer solution, but only allocates costs as a fractional solution, duality implies that price-of-fair-sharing is lower-bounded by the integrality gap of the problem. It is folklore that the price-of-fair-sharing is $\alpha$ whenever the integrality gap of the natural LP relaxation is $\alpha$ \cite{jain_mahdian_2007,meir2010minimal}.

The relationship between the price-of-fair-sharing and the integrality gap of the natural IP/LP formulation does not, however, appear to hold for CIPs.
The multi-cover constraints in CIPs render the natural linear programming relaxation ill-suited for cost-sharing. Even with just one user $|U| = 1$ and integral inputs, the integrality gap of the CIP is unbounded \cite{carr1999strengthening}. Naively, this would imply that the price-of-fair-sharing in CIPs is unbounded. However, Li, Sun, Wang, and Lou \cite{Li2005CovCostShare} show that a bound can be attained for the special case of CIPs with integer-valued $\bfA$ and $\bfr$. Their analysis, however, does not make explicit use of duality. Moreover, our IoT application yields CIPs that do not satisfy the assumption of integer-valued $\bfA$ and $\bfr$. This begs two questions: How do cost-sharing and duality relate in CIPs, and can there be an effective cost-recovery at all when data are non-integer-valued?

\section{Knapsack-Cover Constraints to the Rescue}

This paper presents a principled framework for finding cost-shares in CIPs using linear programming duality. Our approach makes use of a well-known strengthened LP formulation based on knapsack-cover inequalities \cite{carr1999strengthening}. Our main contribution is to show that every feasible dual solution in this strengthened LP, which we shall denote KC-LP, naturally induces cost-shares that satisfy the core property. This has significant algorithmic consequences. First, our results imply that any approximation algorithm that produces CIP solutions with cost at most $\alpha$ times the KC-LP optimum can be used to produce cost-shares that recover $1/\alpha$ of the cost; that is, a price-of-fair-sharing of $\alpha$. There are many such algorithms \cite{carr1999strengthening,chekuri2019approximating,chen2021partial,kolliopoulos2005}. More generally, our framework can be used to find cost-shares for any integer CIP solution by solving the strengthened dual linear program. 

These methods yield the first cost-sharing algorithm with bounded price-of-fair-sharing for general CIPs. Furthermore, we prove that the implied price-of-fair-sharing bounds are tight. In addition, we showcase the efficacy of our framework by recovering up to 93\% of the cost in semi-stylized LoRaWAN covering problems at scale; this is over twice the recovery of the next-best method. We also use
our KC-LP approach
to obtain analogous results for a more general setting in which the service provider also selects a subset of users to receive the service based on the users' private valuations elicited from a mechanism that is group-strategyproof. This reinforces the central message of this paper, affirming the powerful connection between an effective cost sharing mechanism and KC-LP dual solutions.

The knapsack-cover inequalities are introduced in the seminal work by Carr, Fleischer, Leung, and Phillips \cite{carr1999strengthening} to strengthen the LP so as to bound its integrality gap. It is helpful to first understand how the integrality gap of the natural LP relaxation is unbounded. Carr \textit{et al.} provide the following instance: let $R > 0$ be a large integer requirement for a single user, and let there be two facilities; facility $a$ provides $R-1$ units of coverage at near-zero cost, whereas facility $b$ provides $R$ units of coverage at a cost of $1$. Clearly, a feasible integer solution must include item $b$, with a total cost of $1$. A fractional solution, however, can select a full unit of item $a$, and only a $1/R$ fraction of item $b$, with a cost of $1/R$. The resulting integrality gap is $R$, which can be arbitrarily large.

Carr \textit{et al.} \cite{carr1999strengthening} derive a strong bound on the integrality gap by adding 
exponentially many valid \textit{knapsack-cover} (KC) inequalities to produce a strengthened LP relaxation. These inequalities represent residual coverage requirements at partial solutions. Suppose facilities $S \subseteq \facs$ have been built. Now user $j$ has a \emph{residual requirement} $r^S_j = \max\left\{r_j - \sum_{i \in S}a_{ij}, 0\right\}$. To satisfy $j$'s residual requirement, an additional contribution of at least $r^S_j$ is needed from the remaining unbuilt facilities $\facs \backslash S$. In addition, there is no benefit to exceeding $r^S_j$, so the original contributions $a_{ij}$ can be clipped
to $r^S_j$ if they exceed this value. We call $a^S_{ij} = \min\left\{a_{ij}, r^S_j\right\}$ the \textit{residual contribution}, and set it to zero if $i$ is in $S$.
This defines the {\it knapsack-cover inequality}:
\begin{equation}
\label{eq:knapsack-cover}
\sum_{i \in \facs \backslash S}a^S_{ij} \geq r^S_j.
\end{equation}
There are $|\users| \times |2^\facs| = m2^n$ KC inequalities. The KC inequalities are valid; their presence does not change the set of feasible \textit{integer} solutions, and
reduces the feasible set of the LP relaxation. The strengthened linear program is as follows:
\begin{align}
\label{eq:primal_2}
\min_x ~ &\sum_{i \in \facs} c_i x_i, \nonumber \\
s.t. \quad & \sum_{i \in \facs \setminus S} a^S_{ij} x_{i} \geq r^S_j,  &&\forall (j, S)  \in \users \times 2^\facs, \tag{KC-LP}\\
&  x_{i} \geq 0, &&\forall i \in \facs. \nonumber 
\end{align}
Note that the constraints $x_i \leq 1$ have been dropped. (Multiplicity constraints are implicit, in that $a^S_{ij} = 0$ whenever $i \in S$.) Associated with this primal is the knapsack-cover dual program (\ref{eq:dual_2}), referred to as the KC-LP dual. The dual has one constraint for each facility, and one variable for every KC inequality: 
\begin{align}\label{eq:dual_2}
    \max_{y} ~  & \sum_{j \in \users} \sum_{S \subseteq \facs}r^S_j y^S_j, \nonumber\\
    s.t. \quad & \sum_{j \in \users}\sum_{S \subseteq \facs \setminus \{i\}} a^S_{ij}y^S_j \leq c_i, && \forall i \in \facs, \tag{KC-DP} \\
    & y_j^S \geq 0,  && \forall (j, S) \in \users \times 2^{\facs}.\nonumber
\end{align}

There are known approximation algorithms for CIPs that yield integer solutions with costs within a multiplicative factor of the KC-LP optimum. The worst-case performance of these algorithms depend on the column and row sparsity of the contributions $\bfA$. Let $\Delta = \max_i \{\sum_{j \in \users}\mathbf{1}[a_{ij} > 0]\}$ denote the column sparsity, and let $\Gamma = \max_j\left\{ \sum_{i \in \facs}\mathbf{1}[a_{ij} > 0]\right\}$ denote the row sparsity. Carr \textit{et al.} \cite{carr1999strengthening} propose a $\Gamma$-approximation algorithm based on rounding a KC-LP solution; Kolliopoulos and Young \cite{kolliopoulos2005} design an alternative KC-LP rounding procedure to yield a $\bigO{\log(1 + \Delta)}$-approximation algorithm. Together, these provide upper bounds on the strengthened integrality gap of $\Gamma$ and $\bigO{1 + \log\Delta})$. Some algorithms attain improved guarantees at the cost of small violations of the multiplicity constraints \cite{chekuri2019approximating,chen2021partial}. Common to these LP-rounding rounding methods is that they first require obtaining an optimal solution to the KC-LP.

The strengthened linear program and its dual can be solved to near-optimality in polynomial time. Chekuri and Quanrud \cite{chekuri2019approximating} develop a multiplicative weights method that returns approximately optimal primal and dual solutions in near linear time, but with an $\bigO{1/\epsilon^5}$ dependence on the relative error $\epsilon$. Their approach builds upon earlier work of Plotkin, Shmoys, and Tardos \cite{plotkin1995fast}, and the solution method outlined by Carr \textit{et al.} \cite{carr1999strengthening}. The returned dual solutions are feasible, lie within a $\frac{1}{1 - \epsilon}$-factor of the optimum, and have a polynomially-bounded number of non-zero variables \cite{chekuri2019approximating}. In practice, the primal and dual problems can be solved exactly using column generation for quite large instances; the problem of finding a \textit{most violated inequality} can be reduced to solving a sequence of pseudopolynomially-many minimum-cost knapsack problems, each of which admits a pseudopolynomial exact algorithm (e.g., \cite{lawler1977fast}). In our case study, we solve the KC-LP dual to optimality using this approach in large-scale problem instances with thousands of users and thousands of facilities.

\section{Cost-shares and the strengthened LP}

Our main technical contribution is to show that any feasible solution to the strengthened dual (\ref{eq:dual_2}) produces a cost-allocation that 
satisfies the core property. This reconciles the apparent inconsistency for the CIP with respect to the folk understanding  cost-shares and duality.
The cost-shares themselves are remarkably intuitive: each user pays the part of the dual objective associated with her share of service utilization.
\begin{theorem}
\label{thm:dual_cost_shares}
Let $\bfy = (y^S_j)_{S \subseteq \facs, j \in \users}$ be a KC-LP dual-feasible solution. Then, cost-shares
\begin{equation}
\label{eq:cost-shares}
\xi_j := \sum_{S \subseteq \facs}r^S_j y^S_j, \qquad \forall j \in \users,
\end{equation}
satisfy the core property. We say the cost-shares in (\ref{eq:cost-shares}) are \textrm{induced by} $\bfy$.
\end{theorem}
Clearly, by summing the cost-shares over all users, we recover the dual objective in (\ref{eq:dual_2}). The theorem is simple, but its consequences are profound. In particular, it can be used to exploit approximation algorithms that find integer solutions with costs bounded in terms of a corresponding dual-feasible solution. 
\begin{corollary}
     Let $X \subseteq \facs$ be a feasible solution to the CIP, and $\bfy$ a KC-LP dual feasible solution. If $X$ is KC-LP-relative $\alpha$-approximation, i.e.,
   \begin{equation*}
        c(X) \equiv \sum_{i \in X}c_i \leq \alpha \sum_{j \in U}\sum_{S \subseteq \facs}r^S_j y^S_j,
    \end{equation*}
    then the cost-shares induced by $\bfy$ have cost-of-fair-sharing at most $\alpha$ times $c(X)$. 
\end{corollary}
This implies that the Carr \textit{et al.} \cite{carr1999strengthening} algorithm yields cost-allocations with cost-of-fair-sharing at most  $\Gamma \leq m$, and that the rounding approach of Kolliopoulos and Young \cite{kolliopoulos2005} produces a cost-of-fair-sharing of $\bigO{ \ln\Delta}$, where $\Delta \leq n$. Moreover, if small violations of the multiplicity bounds are tolerated, this is improved further on sparse instances \cite{chekuri2019approximating,chen2021partial}. These are the first price-of-fair-sharing bounds for general CIPs. Furthermore, even on the more restricted multi-set multi-cover problem, i.e., CIPs with integer inputs, our work yields improvements. The $\Gamma$-bound is new, and the latter $\bigO{\ln \Delta}$ bound dominates the existing $\mathcal{O}(\max_i \sum_ja_{ij})$ bound of Li \textit{et al.} \cite{Li2005CovCostShare}. Finally, our theorem implies that one can also find cost-shares directly by solving the KC-LP dual. This decouples the problem of finding an integer CIP solution from that of finding cost-shares. Our case study shows this can have great value in practice.

\paragraph{Proof of Theorem 1} The proof is relatively simple; it naturally uses dual feasibility, and a rearrangement of summations that is standard in the analysis of primal-dual schema (see, e.g., \cite{ShmoysWilliamsonDesign}). Fix a KC-LP dual feasible solution $\bfy$, and let $\mathbf{\xi}$ be the induced cost-shares. The main burden of proof is to show that no subset of users has incentive to act separately from the others. In other words, we need to show that for any $J \subseteq \users$,
\begin{equation*}
   c^*_J \geq  \sum_{j \in J}\xi,
\end{equation*}
where $c^*_J$ is the minimum cost of an integer solution serving group $J$ only; this is the minimum cost to the CIP in which only constraints associated with users $J$ are included. 
 
We prove this by considering an optimal solution to the problem of serving the smaller set of users. Let $X^*_J \subseteq \facs$ be a minimum-cost solution for serving group $J$. Using \ref{eq:dual_2} feasibility of $\bfy$,
 we see that
\begin{align*}
    c^*_J &= \sum_{i \in X^*_J}c_i 
    \geq \sum_{i \in X^*_J}\left(\sum_{j \in \users}\sum_{S \subset \facs \backslash \{i\}}a^S_{ij}y^S_j \right) \geq \sum_{i \in X^*_J}\sum_{j \in J}\sum_{S \subset \facs \backslash \{i\}}a^S_{ij}y^S_j, 
\end{align*}
where the last inequality is due to the fact that all variables are non-negative.

The last part can be shown using a standard change in the order of summation. The right-hand-side expression above counts, for every item $i \in X^*_J$, the sum over subsets $S \subseteq \facs$ that do not contain $i$. Equivalently, one can sum, for each subset $S \subseteq \facs$, every item in $X^*_J$ not in $S$, i.e. 
\begin{equation}
\label{eq:proof_step1}
    \sum_{i \in X^*_J}\sum_{j \in J}\sum_{S \subset \facs \backslash \{i\}}a^S_{ij}y^S_j = \sum_{j \in J}\sum_{S \subseteq \facs}y^S_j\left(\sum_{i \in X^*_J \setminus S}a^S_{ij}\right)
\end{equation}
Recall that $X^*_J$ is a feasible solution to the sub-problem of serving users in set $J$ only. The residual demand of each subset $S$ is satisfied by $X^*_J$ for all users in $J$:
\begin{equation}
\label{eq:proof_step2}
    \sum_{i \in X^*_J \backslash S} a^S_{ij} \geq r^S_J, \quad \forall j \in J.
\end{equation}
By applying the inequality (\ref{eq:proof_step2}) to (\ref{eq:proof_step1}), we arrive at the cost allocation to group $J$ under the cost-shares $\mathbf{\xi}$. In summary, we have shown that
\begin{equation}
    \label{eq:core-proof}
    c(X^*_J) \geq \sum_{j \in J}\sum_{S \subseteq F}r^S_jy^S_j = \sum_j \xi_j 
\end{equation}
This proves that each group $J \subseteq \users$ prefers to accept the cost-shares $\mathbf{\xi}$ over forming a group on their own. In other words, any \ref{eq:dual_2} dual solution $\bfy$ produces cost-shares satisfying the core property.\hfill $\square$

A natural question is whether there are cost-allocations, perhaps not derived from KC-LP dual variables, that have
lower worst-case prices-of-fair-sharing.
For CIP the answer is \emph{no}; our bounds are provably tight, at least with respect to parameters $\Gamma$ and $\Delta$. In particular, in the special case in which all contributions and requirements are binary, the KC-LP and its dual reduce to the standard set cover linear programs. Here, the folk theorem applies, and the cost-recovery ratio is provably upper-bounded by the integrality gap \cite{DengAlgoAspectsCore}. For Set Cover, and hence CIP as well, the integrality gap can be as large as $\Gamma$ and $\ln \Delta$ \cite{feige1998threshold,Singh2019}. Hence, the lower bounds of our cost-recovery ratios for CIP are tight.

\section{A case study on LoRaWAN coverage}

We conduct a case study to evaluate the effectiveness of the KC-LP cost-sharing framework on a practical coverage-sharing problem for LoRaWAN. The study is based on a scenario in which the goal is to provide coverage over Brooklyn, NY, a relatively large and densely built urban area. The 
challenge
is to find a gateway placement that provides sufficient coverage throughout the area at minimal cost, and to allocate the cost of the gateways across the users, subject to the core constraint, while recovering as much of the cost as possible. This is the main problem motivating our work. Our results are based on the average performance over a sequence of random problem instances.
Our
results suggest the KC-LP framework performs well in practice; by solving both the IP and KC-LP dual to optimality, we recover on average 93\% of the cost of the gateways.

The set of problem instances is derived from a combination of real-world data and assumptions made in the absence of available data. Each instance is a CIP with random contribution matrix $\bfA$, requirements $\mathbf{r}$, and facilities $\mathbf{f}$. The demand points, or users $\users$, are defined as a regular gird of $7,808$ points over the study-area. Each instance uses a sub-sample of size $2,000$. Facility locations are derived from the corners of real building footprints, and sub-sampled down to $4,380$ per instance. This mimics a practical constraint faced by operators who do not have access to every site due to the high access cost. Next, each facility $i$ is associated with a cost $c_i$ uniformly distributed between $0$ and $1$. For the contribution matrix $\bfA$, we use a distance-based Okumura-Hata model with normal noise to generate random connection qualities $a_{ij}$ between each gateway-demand point pair $(i, j)$ \cite{Hata1980}. Finally, to ensure feasibility, each demand point has a requirement equal to the value of building all gateways, divided by a geometrically distributed random variable, sampled independently for each user. A total of 10 instances are used.

For each instance, we solve both the CIP and the KC-LP dual to optimality.
The CIPs are solved using an off-the-shelf IP solver; the KC-LP dual is solved using the column generation routine, similar to \cite{carr1999strengthening}. This produces the optimal cost-shares, in the sense that the
price-of-fair-sharing
matches the instance-specific integrality gap exactly. To reiterate, the ability to find cost-shares by solving the KC-LP dual is a valuable consequence of our work.
We are not aware of another equally tractable method for finding cost-shares of this quality in practice. We also consider additional benchmark algorithms.

The KC-LP solver is compared against two natural approximation algorithms that produce cost-shares: the \textsc{PrimalDual} algorithm and the \textsc{Greedy} algorithm. The \textsc{PrimalDual} algorithm, or dual-ascent algorithm, incrementally grows both a feasible KC-LP dual solution, and an infeasible integer CIP solution. The algorithm terminates as soon as the integer solution is feasible. The main idea behind this algorithm is standard (see e.g. \cite{ShmoysWilliamsonDesign}), and
analogous to
that of \cite{carnes2008primal} in a multi-user setting. Next, we also employ an existing \textsc{Greedy} algorithm that also produces both an integer solution to the CIP, as well as cost-shares \cite{Li2005CovCostShare}. This algorithm produces cost-shares via dual fitting; it greedily selects gateways to add, and amortizes the per-coverage cost of selected facilities into an infeasible KC-LP dual solution. Finally, the dual variables are scaled down by $\log(n)$ to ensure feasibility. We also introduce an improved variant of this algorithm,
called \textsc{Greedy+}. %
By viewing the greedy-generated cost-shares as KC-LP dual variables, these need to be scaled down by worst-case bound, but only the minimal amount to make them KC-LP dual feasible. 

One caveat of both greedy algorithms is that they only seem to work well on CIPs with integer requirements $\mathbf{r}$ and contributions $\mathbf{A}$ \cite{dobson1982worst}. As such, for these algorithms only, we modify the instances by
multiplying
the inputs $\mathbf{A}$ and $\mathbf{r}$ by a large constant, and then rounding the connections up, and the requirements down, to the nearest integer. This modification never compromises feasibility, and does not increase the minimum cost. However, solutions to the rounded CIPs need not be feasible for the original real-valued instance. This is a potential weakness of using greedy algorithms on real-valued CIPs.

\pgfplotstableread[row sep=\\,col sep=&]{
    interval & IP-OPT & KC-LP & PD-Obj & PD-Rev & Gr-Obj & Gr-Rev & Gr+-Rev \\
    0--6     & 1      & 0.836 & 1.191  & 0.466  & 1.191  & 0.092  & 0.300 \\
    7--13    & 1      & 0.949 & 1.488  & 0.587  & 1.169  & 0.092  & 0.280 \\
    14--20   & 1      & 0.928 & 1.227  & 0.448  & 1.073  & 0.095  & 0.245 \\
    21--27   & 1      & 0.919 & 1.343  & 0.472  & 1.104  & 0.097  & 0.228 \\
    28--34   & 1      & 0.903 & 1.357  & 0.492  & 1.082  & 0.103  & 0.224 \\
    35--41   & 1      & 0.941 & 1.409  & 0.434  & 1.075  & 0.101  & 0.252 \\
    42--48   & 1      & 0.925 & 1.358  & 0.487  & 1.077  & 0.109  & 0.203 \\
    49--55   & 1      & 0.920 & 1.308  & 0.442  & 1.057  & 0.105  & 0.217 \\
    56--62   & 1      & 0.973 & 1.372  & 0.392  & 1.084  & 0.113  & 0.174 \\
    63--69   & 1      & 0.993 & 1.147  & 0.323  & 1.075  & 0.118  & 0.177 \\
    }\mydata

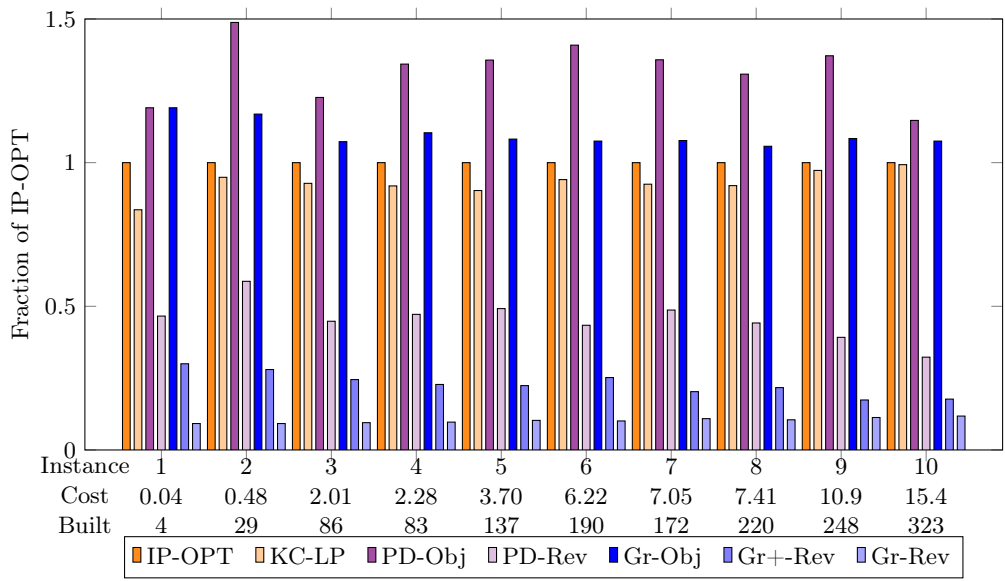
\begin{figure}
\label{fig:bars}
\makebox[\textwidth][c]{
    \begin{tikzpicture}
    \begin{axis}[
            ybar=1.5px, 
            ymin=0, ymax=1.50,
            bar width=2.9px, 
            width=1.0\textwidth, 
            symbolic x coords={0--6,7--13,14--20,21--27,28--34,35--41,42--48,49--55,56--62,63--69},
            legend style={at={(0.5,-0.2*\axisdefaultheight)},
                          anchor=north,legend columns=-1},
            x tick label style={align=center, text height=1.5ex, yshift=-0.11*\axisdefaultheight},
            xticklabels={
                {}, 
                \strut 1 \\ \strut 0.04 \\ \strut 4,
                \strut 2 \\ \strut 0.48 \\ \strut 29,
                \strut 3 \\ \strut 2.01 \\ \strut 86,
                \strut 4 \\ \strut 2.28 \\ \strut 83,
                \strut 5 \\ \strut 3.70 \\ \strut 137,
                \strut 6 \\ \strut 6.22 \\ \strut 190,
                \strut 7 \\ \strut 7.05 \\ \strut 172,
                \strut 8 \\ \strut 7.41 \\ \strut 220,
                \strut 9 \\ \strut 10.9 \\ \strut 248,
                \strut 10 \\ \strut 15.4 \\ \strut 323,
            },
            ylabel={Fraction of IP-OPT},
        ]
        \addplot[fill=orange!90!white] table[x=interval,y=IP-OPT]{\mydata};
        \addplot[fill=orange!40!white] table[x=interval,y=KC-LP]{\mydata};
        \addplot[fill=violet!70!white] table[x=interval,y=PD-Obj]{\mydata};
        \addplot[fill=violet!25!white] table[x=interval,y=PD-Rev]{\mydata};
        \addplot[fill=blue!100!white] table[x=interval,y=Gr-Obj]{\mydata};
        \addplot[fill=blue!50!white] table[x=interval,y=Gr+-Rev]{\mydata};
        \addplot[fill=blue!35!white] table[x=interval,y=Gr-Rev]{\mydata};

        \legend{IP-OPT $\ $, KC-LP $\ $, PD-Obj $\ $ , PD-Rev $\ $, Gr-Obj $\ $, Gr+-Rev $\ $, Gr-Rev}
    \end{axis};
    \node at (-0.0,-0.2) {Instance};
    \node at (-0.0,-0.6) {Cost};
    \node at (-0.0,-1.0) {Built};
    \end{tikzpicture}
}%
\caption{Costs incurred and recovery for 10 instances. IP-OPT is the cost of the IP optimum, KC-LP the cost of the KC-LP optimum. Prefixes PD, Gr, and GR+, represent the \textsc{PrimalDual}, \textsc{Greedy}, and \textsc{Greedy+}, respectively. Suffixes Obj and Rev represent the integer objective cost, and cost-share revenue, respectively.}
\end{figure}

The results are summarized in Figure \ref{fig:bars}. 
Each set of bars along the x-axis represent one instance; the bar heights represent costs.
The \textsc{Greedy} algorithm performs remarkably well, with its solution cost exceeding the minimum by only 10\% on average, whereas the primal-dual algorithm performs worse, averaging nearly 30\% above optimal. 
More remarkably, the \textsc{Optimal} method recovers 93\% of the cost on average; over twice as much as the next best algorithm.
\textsc{Greedy} recovers relatively little cost, even after the improved dual-fitting \textsc{Greedy+}; \textsc{PrimalDual} recovers considerably more -- 45\% of the minimum cost.

Overall, the KC-LP framework finds
near perfect
cost-shares when sharing the cost of LoRaWAN coverage. Using column generation to solve the KC-LP dual, one can recover nearly all of the cost of reasonably realistic coverage provision instances. If the instances are larger such that solving the IP and KC-LP to optimality is prohibitive, one can use both \textsc{Greedy} and \textsc{Primal-Dual}; the former to solve the CIP, the latter to find cost-shares. Alternatively, one can solve the KC-LP to near-optimality using the algorithms methods of \cite{carr1999strengthening,chen2021partial}. The consistency of performance across this data set (along with other results that achieved analogous performance on variants in which the costs were Gaussian rather than uniform) demonstrate the effectiveness of this theoretically-inspired approach to deliver significant results in our application.

\section{Group-strategyproof KC-LP cost-shares}
\label{sec:cross_monot}
So far it has been assumed that the group of users to be served is given; sometimes this choice also falls on the service provider. In an extended setting, the goal of the service provider is to elicit private user preferences and design a mechanism for choosing who to serve, in addition to allocating costs to those served. See Jain and Mahdian \cite{jain_mahdian_2007} for more details on definitions. Assume that each user $j \in \users$ has a private utility $u_j$ for receiving service, and has the option of opting out for a utility of $0$. When utilities are unknown, they must be elicited by a mechanism. This creates a challenge, because users and groups of users may be able to strategically misreport their utilities. A mechanism is said to be \textit{strategyproof} if individual users cannot benefit from misreporting their utility, and \textit{group-strategyproof} if no group of users can benefit by colluding to misreport their utilities. The goal of a mechanism in this setting is to elicit preferences in a group-strategyproof manner, decide who to serve, and allocate-costs in a way that maximizes the cost-recovery ratio.

To find a group-strategyproof mechanism, Moulin and Shenker \cite{moulin2001strategyproof} prove that it suffices to have cross-monotonic cost-allocation mechanism. Cost-shares
$\xi: \users \times 2^\users \rightarrow \mathbf{R}^{|\users|}$
are \textit{cross-monotonic} if the cost allocated to user $j$ does not increase when more users are served:
\begin{equation}
    \label{eq:cross-monotonicity}
    \xi_j ( U) \leq \xi_j ( J), \quad \forall j \in J \subseteq  U \subseteq \users.
\end{equation}
Not all cost-shares that satisfy the core property are cross-monotonic. P{\'a}l and Tardos \cite{PalTardos2003group} provided a general approach for using primal-dual algorithms to find cross-monotonic cost-shares in the facility location problem and the rent-or-buy problem. This approach has been used for many other problems as well \cite{jain2008equitable,konemann2005primal,konemann2008group,LeonardiShaefer2004}. Interestingly, imposing cross-monotonicity usually lowers the best attainable cost-recovery ratio  (or equivalently, increases the price-of-fair-sharing) \cite{PalTardos2003group}. Indeed, Immorlica, Mahdian, and Mirrokni \cite{immorlica2008limitations} upper bounded the cost-recovery ratio for cross-monotonic cost allocations using an elegant probabilistic method. In particular, a cross-monotonic cost allocation for set cover can recover at most $1/\Delta$ of the cost, where $\Delta$ is the size of the largest set. Moreover, even if all elements are covered by at most two sets, the cost recovery ratio is $\bigO{(2 + \epsilon)/ n^{1/3}}$. Li \textit{et al.} \cite{Li2005CovCostShare} describe a cross-monotonic cost allocation for the CIP that recovers $1/(2n)$ of the cost without explicit use of duality; we simplify their proof using the strengthened dual, generalize the algorithm to general CIP, and improve this to $1/(2\Delta)$.

Our strengthened linear programming approach can also be used to find cross-monotonic cost-shares. Our mechanism uses a primal-dual algorithm following the general framework of P{\'a}l and Tardos \cite{PalTardos2003group}. Mechanically, our algorithm is equivalent to that of Li \textit{et al.} \cite{Li2005CovCostShare}, but our analysis is different, in that it uses the KC-LP dual.
\begin{theorem}
    Fix users $U \subseteq \users$. The mechanism (Algorithm \ref{alg:cross-monotone}) produces a feasible solution $X$ for users $U$, and cross-monotone cost-shares $\xi_j ( J)$ that satisfy $\sum_{j \in J}\xi_j ( J) \geq \left(\frac{1}{2\Delta} \right) \cdot c(X)$.
\end{theorem}
The main algorithmic idea behind the mechanism is to let each user independently select facilities in complete isolation from the other users. Cross-monotonicity is enforced by preventing any interactions between users' dual variables. Moreover, the problem faced by an individual user is a minimum  cost-knapsack problem, each of which can be solved by a primal-dual algorithm (Algorithm \ref{alg:primal-dual-ks}) \cite{carnes2008primal}. This produces, for each user $j \in U$, a selection of facilities $X_j$, and a KC-LP dual solution $\bfy_j$ that is feasible for the individual problem with constraints for user $j$ only, and no variables corresponding to other users. Finally, our mechanism selects the union of all selected facilities $(X_j)_{j \in U}$, and scales down the individual dual variables $\bfy_j$ by the column sparsity $\Delta$. The procedure is summarized in Algorithm \ref{alg:cross-monotone}, in which \texttt{MinCostKnapsackPrimalDual} is Algorithm \ref{alg:primal-dual-ks}.

\begin{center}
\begin{minipage}{.8\linewidth}
\begin{algorithm}[H]             
    \DontPrintSemicolon
    \label{alg:cross-monotone}
    \KwIn{$(\facs, \users, \bfr, \bfc, \bfA)$}
    \For{all users $j \in J$ independently}{
    $X_j, \bfy'_j \gets \texttt{MinCostKnapsackPrimalDual}(\facs,r_j, c, a_j)$\\
    $\bfy_j \gets \bfy'_j / \Delta$ 
    }
    $X \gets \cup_{j \in \users}X_j$\\
    $\xi_j ( U) \gets \frac{1}{2\Delta}\sum_{S \subseteq F}y^S_j$ for all $j \in U$\\
    \Return{$X, \xi$}
    \caption{A cross-monotonic primal-dual algorithm}
\end{algorithm}
\end{minipage}
\end{center}

Carnes and Shmoys \cite{carnes2008primal} develop and analyze a primal-dual algorithm for the minimum-cost knapsack problem based on the KC-LP formulation.  Fix a single user $j \in \users$ and let $\bfa_j = (a_{1j}, \dots, a_{nj})$ denote their contributions. The user starts with an all-zero dual solution $\bfy_j$, and an empty selection $X = \emptyset$. While the residual demand $r^X_j$ is positive, they increase the dual variable $y^X_j$. Eventually, some constraint $\sum_{S \subseteq F - \{i\}}a^X_{ij}y^X_j \leq c_i$ becomes tight for a facility $i$. This facility is added to the selection $X$, and the process repeats. This procedure is summarized in Algorithm \ref{alg:primal-dual-ks}. The algorithm returns a selection of facilities $X_j$ and a feasible KC-LP dual solution. Critically, the cost of $X_j$ is at most twice the KC-LP dual objective under $\bfy_j$.
\begin{theorem}[Carnes and Shmoys \cite{carnes2008primal}]
\label{thm:carnes_shmoys}
    Let $X_j \subseteq \facs$ and $\bfy_j$ be a selection of facilities, and the corresponding dual solution returned by Algorithm \ref{alg:primal-dual-ks}. These satisfy the inequality
    \begin{equation*}
        \sum_{i \in X_j}c_i \leq 2 \sum_{S \subseteq \facs}r^S_jy^S_j.
    \end{equation*}
\end{theorem}
This result is used in two parts of our proof. First, we use the dual feasibility of the individual dual variables $\bfy_j$ to construct a feasible dual solution to the master CIP, which gives us the core-property via Theorem \ref{thm:dual_cost_shares}. Secondly, the approximation ratio is used to derive our cost-recovery ratio.

\begin{center}
\begin{minipage}{.8\linewidth}
\begin{algorithm}[H]  
    \label{alg:primal-dual-ks}
    \DontPrintSemicolon
    \KwIn{$(\facs, \bfc, r_j, \bfa_j)$}
    $X, \bfy_j \leftarrow \emptyset, \mathbf{0}$\\
    \While{$r^X_j > 0$}{
    Increase $y^{X}_j$ until for some $i \in \facs \backslash X$:
    \begin{equation}
        \sum_{S \subseteq \facs - \{i\}}a^X_{ij}y^X_j = c_i
    \end{equation}
    $X \leftarrow X \cup \{i\}$\\
    }
    \Return{$X$, $\bfy_j$}
    \caption{Minimum-Cost Knapsack Primal-Dual Algorithm}
\end{algorithm}
\end{minipage}
\end{center}

\paragraph{Proof of Theorem 2.}
To prove this result, we need to argue for cross-monotonicity, the core-property, and cost recovery. We argue for cross-monotonicity first. Clearly, the dual variables $\bfy'_j$ of each user are independent of other users. Meanwhile, the maximum number of users in $U$ served by any facility, $\Delta$, is monotonically increasing in the size of $U$, so the dual variables $\bfy_j = \bfy'_j / \Delta$ are monotonically decreasing in $U$, as are the induced cost-shares.

The core property is easy to prove using dual feasibility and our Theorem \ref{thm:dual_cost_shares}.  Consider some selected facility $i \in X$.
Then,
\begin{equation*}
    \sum_{j \in \users}\sum_{S \subseteq F \backslash \{i\}}a^S_{ij}y^S_j = \frac{1}{\Delta}\sum_{\{j \in \users: a_{ij}>0\}}\left(\sum_{S \subseteq \facs \backslash \{i\}}a^S_{ij}y'^S_j\right) \leq  \frac{1}{\Delta}\sum_{\{j \in J: a_{ij}>0\}} c_i \leq  c_i.
\end{equation*}
The first equality follows from dropping users not served by facility $i$. The second equality uses the definition of $\bfy_j = \bfy'_j / \Delta$. The following inequality uses the fact that the dual variables are individually feasible to the min-cost knapsack LP of each user $j$. The final inequality follows from the definition of $\Delta$. The above shows that dual variables $(\bfy_j)_{j \in \users}$ are feasible for the CIP induced by users $U$, and thus Theorem \ref{thm:dual_cost_shares} implies that the core-property is satisfied by the accompanying cost-shares $(\xi_j ( U))_{j \in U}$.

Finally, the cost-recovery recovery ratio follows from the approximation ratio of the minimum-cost knapsack algorithm. In particular, observe that
\begin{equation*}
    \sum_{i \in X}c_i \leq \sum_{j \in \users}\sum_{i \in X_j}c_i \leq 2\sum_{j \in \users}\sum_{S \subseteq \facs}r^S_j y'^S_j = 2\Delta \sum_{j \in \users}\sum_{S \subseteq \facs}r^S_jy^S_j.
\end{equation*}
The first inequality is obvious; the second follows from applying Theorem \ref{thm:carnes_shmoys} to each user $j$ in $U$ individually. The last equality follows from the definition of $\bfy_j$. This proves that the
cost-of-fair-sharing is at most $2\Delta$ as claimed.
\hfill $\square$

Finally, the proof suggests there is potential for improvement. In fact, whenever the contributions $\bfA$ are binary, the min-cost knapsack algorithm is exact, in which case the cost-recovery ratio is $1/\Delta$ \cite{Li2005CovCostShare}. Moreover, if we know that each selected facility $X$ always selected by at least two users, it also follows that the
cost-of-fair-sharing is a factor $2$ smaller, i.e. $\Delta$.
On the other hand, no group-strategyproof mechanism can recover more than $1/\Delta$ of the cost in general \cite{immorlica2008limitations}. Whether the $2\Delta$ cost-of-fair-sharing is tight 
when contributions are non-binary remains an open problem \cite{Li2005CovCostShare}.



\begin{credits}
\subsubsection{\ackname} This material is based on work supported by the NSF under Grant CNS-1952063.

\subsubsection{\discintname}
The authors have no competing interests to declare that are relevant to the content of this article.
\end{credits}

%
%

%

\bibliographystyle{splncs04}
\bibliography{bibliography}

\newpage
\appendix
\section{LoRaWAN, network planning, and the CIP}
\label{sec:lorawan}

Low-Power Wide-Area Networks (LPWANs) are a popular IoT solution for connecting devices to the internet over long distances using little power. LoRaWAN one of the most widely used LPWAN protocols; see e.g., Alumhaya \textit{et al.} \cite{almuhaya2022survey} for an overview. LoRaWAN is designed for the transmission of small packets of data from battery-driven sensors to wireless receivers called \textit{gateways}. These forward the data to the internet over a broadband backhaul \cite{LoRa-technical}.
LoRaWAN applications include road-surface monitoring, sensing fill-levels of municipal garbage containers, and tracking livestock locations, which are used to, respectively, inform safe traffic routing, optimize garbage pickup, and reduce costs of livestock monitoring, respectively. While  LoRaWAN is the primary motivating application of this work, the covering integer program we study is quite general, and therefore applicable to other technologies, not to mention problems entirely divorced from wireless networks and IoT.

A property of LoRaWAN that is particularly relevant to cost-sharing are its low barriers to entry. Specifically, LoRaWAN uses the ISM band of the wireless spectrum \cite{LoRa-technical}. This is an unlicensed band; anyone can operate a gateway free of charge, without a license or certification. Secondly, LoRaWAN gateways are relatively inexpensive, with some models costing just a few hundred US dollars. While the cost of carrier-grade gateways that support large volumes of traffic can be higher, individuals or small groups of users can serve their own needs at relatively low costs. Finally, organizations such as TTN provide open-source software and free online support for managing the server and software end of LoRaWAN networks, which further increases accessibility.
Altogether, these low barriers give users considerable leverage over the cost of wireless coverage. This is in stark contrast with other wireless technologies and natural monopolies, such as 5G, that require costly licenses and hardware.

Another unique and critical feature of LoRaWAN is the use of redundant gateways for improving the quality of coverage. LoRaWAN is \textit{association-free}, in that a single device does not transmit to any particular receiver \cite{LoRa-technical}. Instead, transmissions are broadcast to all receivers, and demodulated by each one that successfully receivers the packet. Duplicates are pruned in hindsight by the network server. This feature is critical for achieving robustness through redundancy; a transmitted packet is lost only if all receivers fail to demodulate it. Studies on LoRaWAN connectivity find that transmissions often fail at random, but the success rate varies predictably with distance, terrain, and wireless parameters \cite{attia2019loraexperimental,augustin2016loraStudy,petric2016measurements}.
By being covered by multiple gateways, the probability of packet loss can be kept minimal despite individual connections frequently failing. The CIP model accommodates user reception rate requirements, gateway redundancy, and heterogeneity in connection qualities.


\textit{Receptions and the link budget}. There are models for determining whether a transmission is received or not. A key concept used in most such models is \textit{received power}, denoted $P_{tx}$. LoRaWaN receivers are said to successfully process any arriving transmission with a received power of at least $p_{min} = -120$ dBm \cite{LoRa-technical}.  Received power is usually modeled using a \textit{link budget}. In simplest form the budget is
\begin{equation}
   \label{eq:link-budget}
   P_{rx} = p_{tx} - L
\end{equation}
Here $p_{tx}$ stands for transmitted power; this represents roughly how ``loud'' a device is ``speaking''. A higher power exhausts batteries faster. Usual for IoT devices is a transmitted power of $10$ dBm, $30$ dBm at most. The variable $L$ stands for \textit{path loss}, which describes how the transmitted power deteriorates as the transmission propagates through space. Modeling the path loss is key to understanding wireless connectivity.

\textit{Hata path loss}. A widely used model for path loss is the Hata \cite{Hata1980} model. It expresses the path loss $L(\bfX)$ over a given wireless link as a simple log-polynomial model over data. The data $\bfX$ are the distance $d$, height of the (mobile) transmitter $h_M$, the height of the receiver (base station) $h_B$, and the frequency $f$, i.e. $\bfX = (d, h_M, h_B, f)$. The Hata path loss is
\begin{align} 
\label{eq:hata-loss}
L(d, h_M, h_B, f) = 69.55 &+ 26.16 \log_{10}(f) - 13.82\log_{10}(h_B) \nonumber\\
&+ (44.9 - 6.55 \log_{10}(h_B))\log_{10}(d) + C_H  
\end{align}
Where $C_H$ is a height-correction factor which in cities, and at $f = 916$ Mhz, is
\begin{equation*}
  C_H =  3.2\left(\log_10(11.75 h_M) \right)^2 - 4.97.
\end{equation*}
It is sometimes further assumed that there is an additive random normal component $X \sim \mathcal{N}(0, \sigma^2)$ in the path loss. This is called \textit{log-normal shadowing}. While these models are simple, we do not take them at face value. We retain our beliefs that there is considerable uncertainty around connectivity.

The additive coverage constraints in the CIP can capture requirements in terms of packet reception probabilities via a simple reduction. Assume that the successful reception from user $j$ to facility $i$ is a binary variable, with given success, or reception, rate $\rho_{ij}$. In this case, one can view coverage provision as a fuzzy set cover problem. Chian, Hwang, and Liu \cite{chiang2005alternative} reduce fuzzy set covering to a CIP. Assume that failures of links are independent, and let $\bfx  = (x_1, \dots, x_n)$ be a binary vector encoding a subset of built facilities $F \subseteq \facs$. Then the failure rate of a transmission from user $j$ is
\begin{equation*}
    \Pr[\text{Failure} \mid F] = \prod_{i \in \facs} (1- \rho_{ij})^{x_i}.
\end{equation*}
Taking logarithms this yields an expression that is additive in the variables $x_i$.
\begin{equation*}
    \log \Pr[\text{Failure} \mid x] =  \sum_{i \in \facs}\log(1 -\rho_{ij})x_i
\end{equation*}
A constraint limiting the maximum packet error rate can therefore be expressed as a covering constraint in the CIP. While independence is a strong assumption, this reduction adds intuition to the CIP, and can be viewed as a stylized approximation to a dependent system, e.g., by choosing more ``pessimistic'' estimates for $\rho_{ij}$, or higher service requirements $r_j$.

\end{document}